\documentclass{article}
\usepackage{spconf,amsmath,graphicx}
\usepackage{algorithm}
\usepackage{algpseudocode}
\usepackage{booktabs}
\usepackage{amsfonts}
\usepackage{hyperref}

\title{Speech enhancement with weakly labelled data from AudioSet}
\name{Qiuqiang Kong, Haohe Liu, Xingjian Du, Li Chen, Rui Xia, Yuxuan Wang}

\address{ByteDance, Shanghai, China \\ \{kongqiuqiang, liuhaohe.0379, duxingjian.real, chenli.cloud, \\ rui.xia, wangyuxuan.11\}@bytedance.com}

\begin{document}
%
\maketitle

\begin{abstract}
Speech enhancement is a task to improve the intelligibility and perceptual quality of degraded speech signal. Recently, neural networks based methods have been applied to speech enhancement. However, many neural network based methods require noisy and clean speech pairs for training. We propose a speech enhancement framework that can be trained with large-scale weakly labelled AudioSet dataset. Weakly labelled data only contain audio tags of audio clips, but not the onset or offset times of speech. We first apply pretrained audio neural networks (PANNs) to detect anchor segments that contain speech or sound events in audio clips. Then, we randomly mix two detected anchor segments containing speech and sound events as a mixture, and build a conditional source separation network using PANNs predictions as soft conditions for speech enhancement. In inference, we input a noisy speech signal with the one-hot encoding of ``Speech'' as a condition to the trained system to predict enhanced speech. Our system achieves a PESQ of 2.28 and an SSNR of 8.75 dB on the VoiceBank-DEMAND dataset, outperforming the previous SEGAN system of 2.16 and 7.73 dB respectively.
\end{abstract}
\section{Introduction}
\label{sec:intro}
Speech enhancement (SE) is a task to improve the intelligibility and perceptual quality of degraded speech signal. Speech enhancement has many applications in our life, such as teleconference, mobile phone calls, automatic speech recognition and hearing aids \cite{loizou2013speech}. Early works of speech enhancement applied signal processing methods such as minimum-mean square error short-time spectral amplitude estimator \cite{ephraim1984speech} and non-negative matrix factorization (NMF) \cite{mohammadiha2013supervised}. Those conventional methods perform well under stationary noise, but have limited performance under non-stationary noise or in low signal-to-noise ratio (SNR) environments. Recently, neural network based methods have been proposed for speech enhancement, such as denoising autoencoder \cite{lu2013speech}, fully connected neural networks \cite{xu2014regression}, recurrent neural networks (RNNs) \cite{weninger2015speech}, convolutional neural networks \cite{park2016fully, fu2016snr} (CNNs) or time domain CNNs \cite{luo2019conv, rethage2018wavenet, pandey2019tcnn} and generative adversarial networks (GANs) \cite{pascual2017segan, donahue2018exploring}. Those neural network based speech enhancement methods require clean speech and background noise for training. Recently, universal source separation systems \cite{wisdom2020unsupervised1, wisdom2020unsupervised2} have been proposed for source separation without clean training data.

However, previous neural network based speech enhancement methods require clean speech and background noise for training, while collecting clean speech and background noise can be difficult and time consuming. For example, the background noises recorded in the laboratories \cite{thiemann2013demand} can be different from real world sounds. It is difficult to collect a large-scale dataset covering a wide range of sounds in our world. In addition, speech datasets such as TIMIT \cite{garofolo1993timit} and VoiceBank \cite{veaux2013voice} contain neutral emotion speech, while there can be various emotions of speech in our real life. Recently, a large-scale AudioSet \cite{gemmeke2017audio} dataset containing hundreds of different sound classes from YouTube was released, which provides a larger variety of sounds than previous speech and noise datasets.

However, the difficulty of using AudioSet for speech enhancement is that audio clips in AudioSet are weakly labelled. That is, each audio clip is only labelled the presence or absence of sound events, without knowing their onset and offset times. Also, AudioSet does not indicate clean speech in audio clips, and speech are usually mixed with other sound events. In this article, we propose a speech enhancement framework trained with weakly labelled data. First, we apply pretrained audio neural networks (PANNs) \cite{kong2019panns} to select 2-second \textit{anchor segments} that are most likely to contain speech or sound events in an audio clip. One contribution of this work is that we propose an anchor segment mining algorithm to better select anchor segments for creating mixtures. Two randomly selected anchor segments are used to constitute a mixture. Then a convolutional UNet \cite{jansson2017singing} is used to predict the waveform of individual anchor segments. We extend the loss function calculated on spectrogram \cite{kong2020source} to a loss function calculated in the waveform domain. For the speech enhancement task, we evaluate various metrics including PESQ, CSIG, etc. that were not discussed in \cite{kong2020source}.

This paper is organized as follows: Section \ref{section:se_wld} introduces our speech enhancement system trained with weakly labelled data. Section \ref{section:experiments} shows the experiment results. Section \ref{section:conclusion} concludes this work. 

\section{Speech Enhancement with Weakly Labelled Data}\label{section:se_wld}

\subsection{Neural Network Based Speech Enhancement}
Recently, neural network based methods have been applied to speech enhancement, and have outperformed conventional speech enhancement methods \cite{xu2014regression}. The neural network based speech enhancement methods require pairs of noisy speech and clean speech for training. We denote a noisy speech as $ x \in \mathbb{R}^{L} $, and its corresponding clean speech as $ s \in \mathbb{R}^{L} $, where $ L $ is the number of samples in an audio clip. Then, a neural network learns a mapping: $ f: x \mapsto s $, where $ f $ can be modeled by a neural network with learnable parameters, such as fully connected neural networks \cite{xu2014regression}, RNNs \cite{weninger2015speech}, CNNs \cite{park2016fully, fu2016snr} and time domain CNNs \cite{luo2019conv, rethage2018wavenet, pandey2019tcnn}. we denote the enhanced speech as $ \hat{s} = f(x) $. In training, the parameters of $ f $ can be optimized by minimizing a loss function $ l(\hat{s}, s) $, such as a mean absolute error (MAE) loss:

\begin{equation} \label{eq:loss}
l_{\text{MAE}} = \| \hat{s} - s \|_{1},
\end{equation}

\noindent where $ \| \cdot \|_{1} $ is an $ l_{1} $ norm. In inference, the enhanced speech $ \hat{s} $ can be calculated by $ \hat{s} = f(x) $. However, one disadvantage of the above neural network based speech enhancement method is that noisy and clean speech pairs are required for training, which can be difficult and time consuming to obtain. To address this problem, we propose a speech enhancement framework that can be trained with weakly labelled data. That is, training a speech enhancement system from audio clips containing noisy speech.

\subsection{Speech Enhancement with Weakly Labelled Data}
Our speech enhancement system is trained with a large-scale weakly labelled AudioSet \cite{gemmeke2017audio} dataset containing 527 kinds of sound classes. Most of audio clips have durations of 10 seconds. AudioSet is weakly labelled, that is, each audio clip is only labelled with tags, but without onset and offset times of sound events. Also, AudioSet does not indicate clean speech, where speech are usually mixed with other sounds under unknown SNR. Previous works has investigated general source separation with weakly labelled data \cite{kong2020source}. Our improvement to \cite{kong2020source} is that we propose a novel anchor segment mining algorithm in Section \ref{section:mining}. To begin with, we denote two anchor segments containing different sound events as $ s_{1} $ and $ s_{2} $ respectively. The anchor segments $ s_{1} $ and $ s_{2} $ are selected from two audio clips that are most likely to contain speech or sound events. The anchor segments $ s_{1} $ and $ s_{2} $ are selected to have disjoint audio tags that is described in Section \ref{section:mining}. In training, we build a neural network to learn a mapping:

\begin{equation} \label{eq:proposed_se}
f(s_{1} + s_{2}, c_{1}) \mapsto s_{1},
\end{equation}

\noindent where $ c_{1} \in [0, 1]^{K} $ is a conditional vector that controls what source to separate, and $ K $ is the number of sound classes in AudioSet. In training, there is no need for $ s_{1} $ or $ s_{2} $ to be clean. The conditional vector $ c_{1} $ is the audio tagging probability calculated on $ s_{1} $. To explain, if $ s_{1} $ contains both ``Speech'' and ``Water''. When conditioning on the audio tagging probability $ c_{1} $, the system (\ref{eq:proposed_se}) will separate both ``Speech'' and ``Water''. In inference, the enhanced speech $ \hat{s} $ can be obtained by input a noisy speech $ x $ and setting the conditional vector $ c $ as the one-hot encoding of ``Speech'':

\begin{equation} \label{eq:inference}
\hat{s} = f(x, c).
\end{equation}

\noindent To explain, the training of the speech enhancement system described in (\ref{eq:proposed_se}) does not require clean speech. Still, we can obtain clean speech from noisy speech using the trained speech enhancement system in (\ref{eq:inference}).

\subsection{Sound Event Detection for Selecting Anchor Segments}\label{section:sed}

The anchor segments $ s_{1} $ and $ s_{2} $ are 2-second segments used to constitute a mixture as input. To begin with, we randomly select two sound classes from AudioSet. For each sound class, we randomly select an audio clip in AudioSet. However, there are no information of when the sound classes occur in audio clips. Therefore, we apply a sound event detection (SED) system \cite{kong2019panns} to predict the frame-wise presence probability of the sound class. The SED system is a DecisionLevelMax system from PANNs \cite{kong2019panns}, which applies log mel spectrogram as input feature, and uses a 14-layer CNN as a classification model. Each convolutional layer has a kernel size of $ 3 \times 3 $. The convolutional layers are followed by a time distributed fully connected layer with $ K $ outputs to predict the frame-wise presence probability of sound classes. The frame-wise predictions are max pooled along the time axis to obtain clip-wise predictions. We denote the weak labels of an audio clip as $ y \in \{0, 1\}^{K} $, and its clip-wise prediction as $ \hat{y} \in [0, 1]^{K} $. The SED system is trained by minimizing a binary crossentropy loss \cite{kong2019panns} between predicted and target weak label tags:

\begin{equation} \label{eq:bce_loss}
\text{loss} = - \sum_{k=1}^{K}y_{k} \text{ln} \hat{y}_{k} + (1 - y_{k}) \text{ln} (1 - \hat{y}_{k}).
\end{equation}

\begin{figure}[t]
  \centering
  \centerline{\includegraphics[width=\columnwidth]{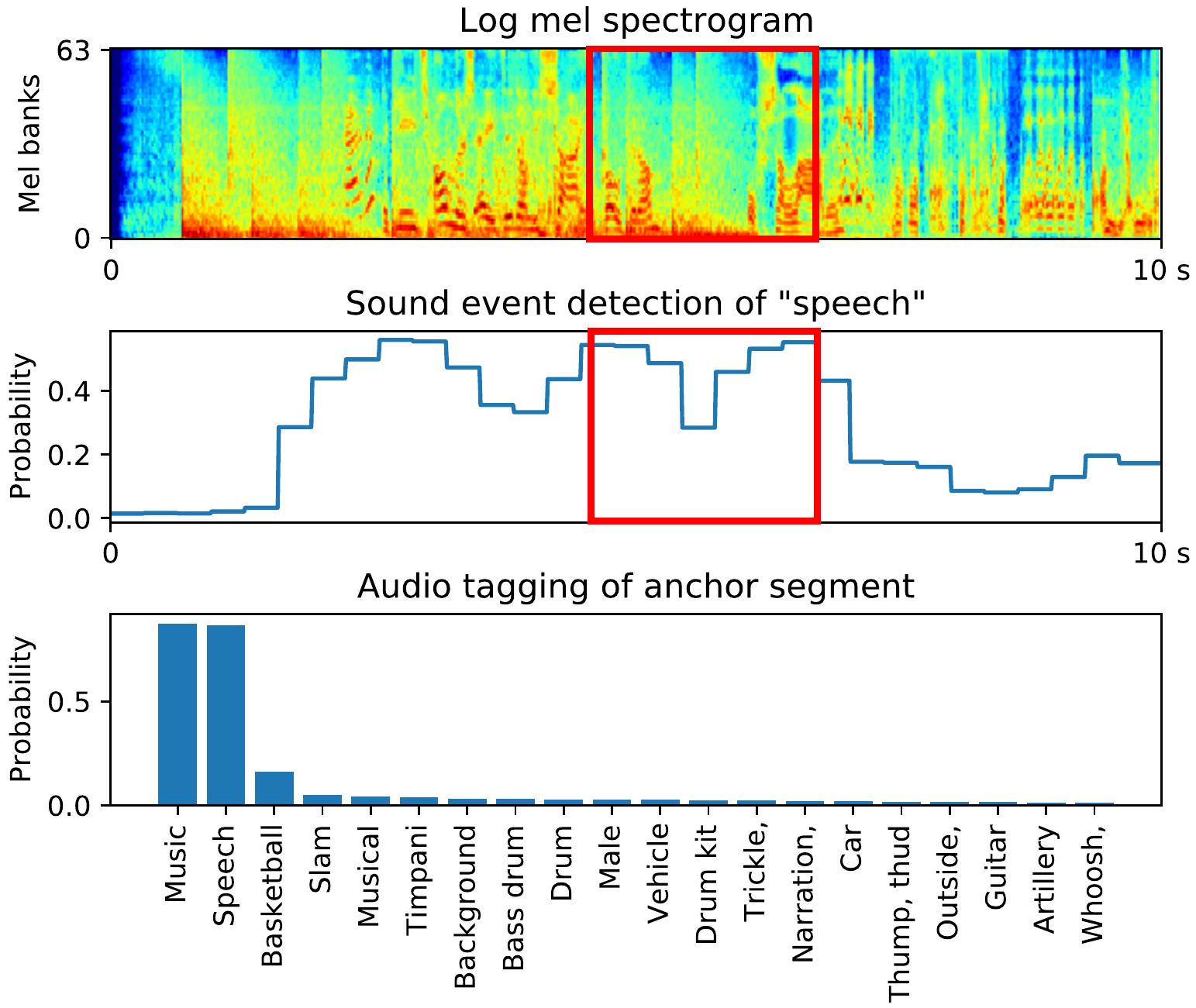}}
  \caption{Top: log mel spectrogram of a 10-second audio clip from AudioSet; Middle: predicted SED probability of ``Speech'', where red block shows the selected anchor segment; Bottom: predicted audio tagging probabilities of the anchor segment.}
  \label{fig:audioset_panns}
\end{figure}

\noindent The first row of Fig. \ref{fig:audioset_panns} shows the log mel spectrogram of a 10-second audio clip from AudioSet containing ``Speech'' and other sound classes. The second row shows the frame-wise SED prediction of ``Speech''. We select anchor segment $ s_{1} $ that is most likely to contain the selected sound event, as shown in the red block in Fig. \ref{fig:audioset_panns}. Similarly, we select anchor segment $ s_{2} $ from another audio clip. Then, we mix $ s_{1} + s_{2} $ as input to (\ref{eq:proposed_se}).

\subsection{Audio Tagging for Constructing Conditional Vector}\label{section:at}
The conditional vector $ c_{1} $ controls what sources to separate from $ s_{1} + s_{2} $. However, there is no ground truth label of $ s_{1} $, so $ c_{1} $ is unknown. In addition, there can be multiple sound events in $ s_{1} $. We apply an audio tagging system on $ s_{1} $: $ c_{1} = g_{\text{AT}}(s_{1}) $ to estimate the conditional vector $ c_{1} $. The audio tagging system $ g_{\text{AT}} $ is a 14-layer CNN of PANNs \cite{kong2019panns}. The 14-layer CNN consists of several convolutional layers. Then, global average and max pooling are applied to summarize the feature maps into a fixed dimension embedding vector. Finally, a fully connected layer is applied on the embedding vector to predict the presence probability of sound events. The training of the audio tagging system applies the binary crossentropy described in (\ref{eq:bce_loss}). The advantage of using audio tagging prediction rather than one-hot encoding of labels to build $ c_{1} $ is that $ g_{\text{AT}}(s_{1}) $ provides a better estimation of sound events probability in $ s_{1} $ than labels. From the top to bottom in Fig. \ref{fig:audioset_panns} shows the log mel spectrogram of a 10-second audio clip, the SED result of the audio clip, and the audio tagging probabilites of the selected anchor segment. For example, the predominant sound events in $ s_{1} $ are ``Music'' and ``Speech''. Other sound classes in $ s_{1} $ include ``Basketball'' and ``Slam'', etc.

\subsection{Anchor Segment Mining}\label{section:mining}
Previous work \cite{kong2020source} proposed to select $ s_{1} $ and $ s_{2} $ randomly from AudioSet. However, if $ s_{1} $ and $ s_{2} $ contain mutual sound classes, the separation result of equation (\ref{eq:proposed_se}) can be incorrect. For example, if both $ s_{1} $ and $ s_{2} $ contain clean ``Speech'', when the conditional vector $ c_{1} $ is the one-hot encoding of ``Speech'', the system in (\ref{eq:proposed_se}) will learn to only separate ``Speech'' from $ s_{1} $, but not ``Speech'' from $ s_{2} $. For a speech enhancement system, we aim to separate all ``Speech'' from both $ s_{1} $ and $ s_{2} $. To address this problem, we propose an anchor segment mining method to select $ s_{1} $ and $ s_{2} $ to have disjoint conditional vectors. In training, we randomly select $ B $ anchor segments to constitute a mini-batch $ \{s_{1}, ..., s_{B}\} $, where $ B $ is the mini-batch size. Then, we calculate the conditional vectors $ \{c_{1}, ..., c_{B}\} $ by $ g_{\text{AT}}(s_{1}) $. For each conditional vector $ c_{b} $, we apply thresholds to predict their present tags $ r_{b} $, where the thresholds are calculated from PANNs \cite{kong2019panns} with equal precision and recall for sound classes. Then, we propose a mining algorithm described in Algorithm \ref{alg:mining} to select pairs of anchor segments to have disjoint predicted tags from the mini-batch to constitute $ s_{1} $ and $ s_{2} $.

\begin{algorithm}[t]
	\caption{Anchor segment mining.}\label{alg:mining}
	\begin{algorithmic}[1]
	    \State Mini-batch of anchor segments: $ S = \{ s_{1}, ..., s_{B} \} $, and their predicted tags: $ R = \{r_{1}, ..., r_{B}\} $.
	    \For {$ r_{1} \in R $}
	        \For {$ r_2{} \in R $}
	            \If{$ r_{1} \cap r_{2} = \o $}
	                \State Collect anchor segments of $ r_{1} $ and $ r_{2} $ to constitute $ s_{1} $ and $ s_{2} $.
	                \State Remove $ r_{1} $ and $ r_{2} $ from $ R $.
	            \EndIf
	        \EndFor
	    \EndFor
	        
	\end{algorithmic}
\end{algorithm}

\subsection{Separation Model}
We apply convolutional UNets \cite{jansson2017singing, kong2020source} on the spectrogram of mixture to build separation systems. To begin with, the waveform of a mixture is transformed into a spectrogram. A UNet consists of an encoder and a decoder. The encoder consists of 12 convolutional layers with kernel sizes of $ 3 \times 3 $ to extract high-level representations. Downsampling layers with sizes of $ 2 \times 2 $ are applied to every two convolutional layers. The decoder is symmetric to the encoder with 12 convolutional layers. Transposed convolutional layers are used to upsample feature maps after every two convolutional layers. Shortcut connections are added between encoder and decoder layers with same hierarchies. In each convolutional layer, the conditional vector $ c_{1} $ is multiplied with a learnable matrix, and is added to the feature maps as biases. This bias information controls what sound events to separate from a mixture. The decoder outputs a spectrogram mask with values between 0 and 1, and is multiplied to the mixture spectrogram to obtain the separated spectrogram of $ s_{1} $. Then, an inverse short time Fourier transform (ISTFT) is applied on the separated spectrogram using the phase of mixture to obtain $ \hat{s}_{1} $. The separation system is trained by minimizing the loss function (\ref{eq:loss}).

\section{Experiments}\label{section:experiments}
Our speech enhancement system is trained on the balanced subset of the weakly labelled AudioSet \cite{gemmeke2017audio} containing 20,550 audio clips with 527 sound classes. The audio clips have durations of 10-second. Audio clips are weakly labelled, and there can be multiple sound events in an audio clip. There are 5,251 audio clips containing ``Speech''. To begin with, we resample all audio clips to 32 kHz to be consistent with the configuration of PANNs \cite{kong2019panns}. The sound event detection and audio tagging systems from PANNs are used to select anchor segments as described in Section \ref{section:mining}. To build the separation system, we extract spectrograms of mixtures using short time Fourier transform (STFT) with a window size 1024 and a hop size 320. All anchor segments have durations of 2 seconds. We set mini-batch size to 24. Adam optimizer \cite{kingma2014adam} is used for training. We trained the system for 1 million iterations using a single Tesla-V100-SXM2-32GB GPU card in one week.

We evaluate our proposed speech enhancement system directly on the test set of the VoiceBank \cite{veaux2013voice} and DEMAND \cite{thiemann2013demand} datasets without training on them. There are 824 paired noisy and clean speech for testing in VoiceBank-DEMAND. Each audio clip has a sample rate of 48 kHz. The noisy speech have four SDR settings of 15, 10, 5 and 0 dB. There are 10 types of noise, including 2 types of synthetic noise and 8 types of noise from DEMAND. There are 28 speakers from VoiceBank. The major difference between our speech enhancement method with previous works is that, we do not use the training data from VoiceBANK-DEMAND, and directly evaluate our speech enhancement system on the test clips.

\begin{figure}[t]
  \centering
  \centerline{\includegraphics[width=\columnwidth]{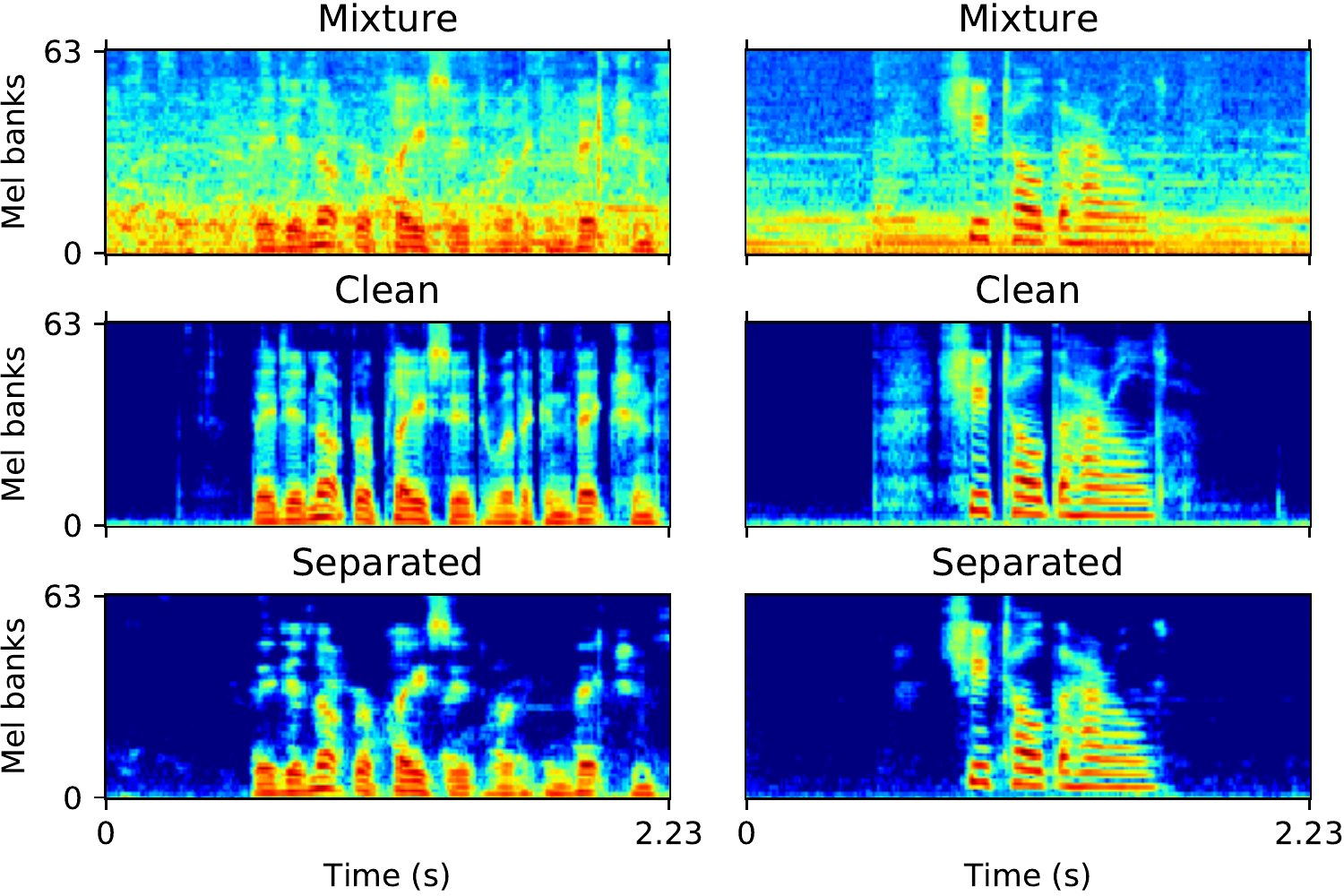}}
  \caption{The left and right columns show two examples of  speech enhancement. Top: log mel spectrogram of noisy speech; Middle: ground truth clean speech; Bottom: enhanced speech.}
  \label{fig:results}
\end{figure}

Following previous works of speech enhancement \cite{scalart1996speech, pascual2017segan, macartney2018improved}, we apply Perceptual evaluation of speech quality (PESQ) \cite{recommendation2001perceptual}, Mean opinion score (MOS) predictor of signal distortion (CSIG), MOS predictor of background-noise intrusiveness (CBAK), MOS predictor of overall signal quality (COVL) \cite{hu2007evaluation} and segmental signal-to-ratio noise (SSNR) \cite{quackenbush1988objective} to evaluate the speech enhancement performance. Table \ref{table:results} shows 
that noisy speech without enhancement achieves PESQ, CSIG, CBAK, COVL, SSNR of 1.97, 3.35, 2.44, 2.63 and 1.68 dB respectively. Our proposed speech enhancement system achieves a PESQ of 2.28, outperforming the Wiener \cite{scalart1996speech} and SEGAN \cite{pascual2017segan} systems. Our system achieves a CBAK of 2.96 and an SSNR of 8.75 dB, outperforming the Wiener and SEGAN systems of 2.68 and 5.07 dB, indicating the effectiveness of training speech enhancement with weakly labelled data. On the other hand, our system achieves a CSIG of 2.43 and COVL of 2.30, lower than other systems, indicating that our speech enhancement may lost details of speech, especially the high frequency component shown in Fig. \ref{fig:results}. The left and right columns of Fig. \ref{fig:results} visualizes two speech enhancement examples of our proposed system. From top to bottom rows show the log mel spectrogram of noisy speeches, target clean speeches and enhanced speeches respectively. Considering that our system is trained with weakly labelled data only, and does not use any training data from VoiceBank-DEMAND. We show that training a speech enhancement system from weakly labelled data is possible. We provide our speech enhancement demos in the following links\footnote{\url{https://www.youtube.com/watch?v=q3hVnpNcpBI}}\footnote{\url{https://www.youtube.com/watch?v=DzQvn820u8E}}.

\begin{table}[t]
\centering
\caption{Speech enhancement results}
\label{table:results}
\resizebox{\columnwidth}{!}{%
\begin{tabular}{*{6}{c}}
 \toprule
 & PESQ & CSIG & CBAK & COVL & SSNR \\
 \midrule
 Noisy & 1.97 & 3.35 & 2.44 & 2.63 & 1.68 \\
 Wiener \cite{scalart1996speech} & 2.22 & 3.23 & 2.68 & 2.67 & 5.07 \\
 SEGAN \cite{pascual2017segan} & 2.16 & 3.48 & 2.94 & 2.80 & 7.73 \\
 Wave-U-Net \cite{macartney2018improved} & 2.40 & 3.52 & 3.24 & 2.96 & 9.97 \\
 \midrule
Proposed & 2.28 & 2.43 & 2.96 & 2.30 & 8.75 \\
 \bottomrule
\end{tabular}}
\end{table}

\section{Conclusion}\label{section:conclusion}
In this work, we propose a speech enhancement system trained with weakly labelled data from AudioSet. Our system does not require clean speech and background noise to train the speech enhancement system. We propose to use the sound event detection and audio tagging system from pretrained audio neural networks (PANNs), and an anchor segment mining algorithm for selecting anchor segments. We build conditional UNet sound separation systems for speech enhancement. Our proposed systems outperform the Wiener and SEGAN systems evaluated with the VoiceBank-DEMAND dataset in the PESQ, CBAK and SSNR metrics without using any training data from VoiceBank-DEMAND. In future, we will continue to investigate general source separation with weakly labelled data.

\small
\bibliography{refs}
\bibliographystyle{IEEEbib}

\end{document}